\documentclass[aps,prl,reprint,showpacs,amsfonts,amsmath,superscriptaddress,floatfix]{revtex4-1}

\usepackage[dvipdfm]{graphicx}
\usepackage{amsmath}

\newcommand{\affA}{%
\affiliation{
     National Institute of Information and Communications Technology\\
     4-2-1 Nukui-kitamachi, Koganei, Tokyo 184-8795, Japan}
     }
\newcommand{\affB}{%
\affiliation{
     Japan Science and Technology Agency, NEC Tsukuba Laboratories
    34 Miyukigaoka, Tsukuba, Ibaraki 305-8501, Japan}
	}

\newcommand{\affC}{%
\affiliation{
     National Institute of Advanced Industrial Science and Technology,
     1-1-1 Umezono, Tsukuba, Ibaraki 305-8568, Japan}
     }

\newcommand{\affD}{%
\affiliation{
    Institute of Quantum Science, Nihon University,
    1-8 Kanda-Surugadai, Chiyoda, Tokyo 101-8308, Japan}
    }

\begin{document}
\title{Quantum receiver beyond the standard quantum limit
of coherent optical communication}

\date{\today}

\author{Kenji Tsujino}
\affA
\affB

\author{Daiji Fukuda}
\affC

\author{Go Fujii}
\affC
\affD

\author{Shuichiro Inoue}
\affD

\author{Mikio Fujiwara}
\author{Masahiro Takeoka}
\author{Masahide Sasaki}
\affA



\begin{abstract}
The most efficient modern optical communication is known as
coherent communication
and its standard quantum limit (SQL) is almost reachable with current technology.
Though it has been predicted for a long time that
this SQL could be overcome via quantum mechanically optimized receivers,
such a performance has not been experimentally realized so far.
Here we demonstrate the first unconditional evidence surpassing
the SQL of coherent optical communication.
We implement a quantum receiver with a simple linear optics configuration
and achieve more than 90\% of the total detection efficiency of the system.
Such an efficient quantum receiver will provide a new way
of extending the distance of amplification-free channels,
as well as of realizing quantum information protocols based
on coherent states and the loophole-free test of quantum mechanics.
\end{abstract}

\maketitle

Coherent communication systems achieve
the best signal-to-noise ratio (SNR)
in conventional optical communications based on laser light.
It consists of coherent-state carriers with
quadrature amplitude modulation (QAM)
and a coherent receiver based on homodyne detections
\cite{PMOCS}.
QAM provides the largest signal distances in phase-space
under the power constraint while a coherent receiver detects them
at the standard quantum limited (SQL) sensitivity.
The SQL in optical communication is naturally defined as
the lowest average error probability obtainable by
directly measuring the modulated physical  observable of coherent states.
For example,
an ideal photon counter
reaches the SQL for
intensity modulation (IM) signals
(often called the shot noise limit: SNL)
while
an ideal coherent receiver
reaches the SQL for QAMs.
However this is not the fundamental quantum limit of optical communication.

\begin{figure}[tb]
\begin{center}
 \includegraphics[width=1\linewidth]{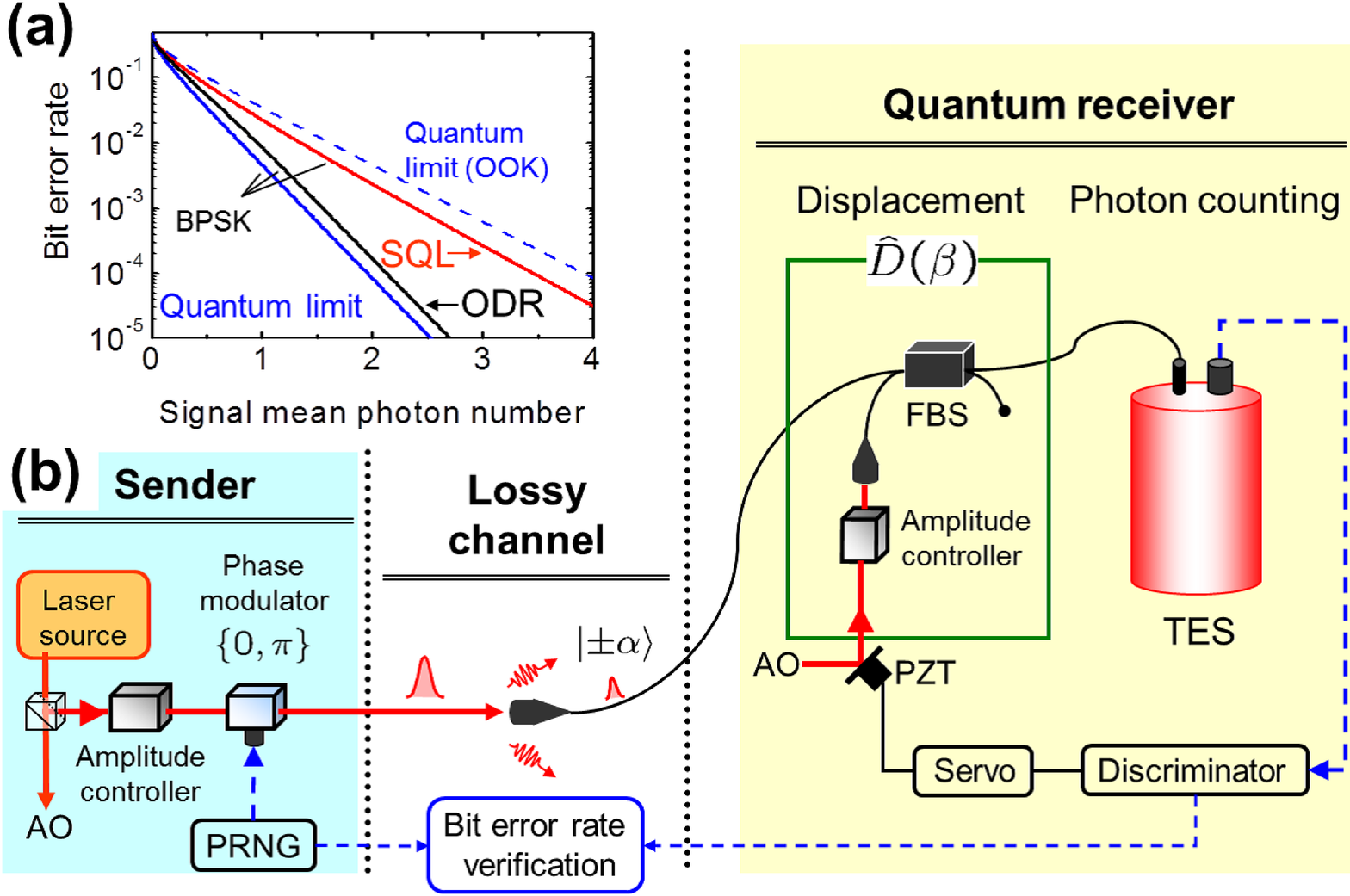}
\end{center}
 \caption{ (color online).
 (a) Bit error rates for the BPSK coherent signals
at the quantum limit (blue solid),
the standard quantum limit (SQL, red solid),
detected by the optimal displacement receiver (ODR, black solid).
As a reference, the blue dashed line is the quantum limit
for an on-off keyed (OOK) signal.
The vertical axis corresponds to the signal mean photon number
reached to the receiver.
 (b) Schematic of a lossy optical channel and
the quantum receiver based on the optimal displacement and
photon counting.
 (c) Experimental setup.
FBS: fiber beam splitter,
AO: auxiliary oscillator,
PRNG: pseudo random-number generator,
TES: transition-edge sensor.
PZT: piezo electric transducer,
}
 \label{fig:schematic}
\end{figure}

One approach to overcome such a limit
is to use eigenstates of the observables, such as
photon number state for IM or (infinitely) squeezed state for QAM,
as carriers \cite{Caves1994}.
For noiseless channels, these non-classical carriers
could completely circumvent the errors due to quantum noise
and in principle realize an error-free communication.
In practice, however, these states are
extremely fragile to losses and easily turned out to be noisy mixed states,
and thus will not work in realistic channels.
Another approach is to keep using coherent states as carriers but
optimize the measurement process.
Coherent state is robust against the linear loss (which is
inevitable in optical channels)
and does not lose its coherence.
Quantum detection theory has predicted that the minimum error bound
is exponentially smaller than the SQL (Fig.~\ref{fig:schematic}(a))
\cite{QDET} and
its implementation schemes have also been proposed
\cite{Kennedy1973,Dolinar1973,Sasaki1996,Takeoka2006,Takeoka2008}.
However, though some proof-of-principle demonstrations have been reported
\cite{Cook2007,Wittmann2008,Tsujino2010}, no unconditional experimental evidence
surpassing the SQL of coherent communication has been reported yet.

In this paper, we demonstrate for the first time,
a quantum receiver outperforming the SQL of
coherent optical communication, i.e. the limit of current optical
communications, in a lossy channel.
We consider the simplest QAM signal set,
binary phase-shift keyed (BPSK) coherent states
$\{|\alpha\rangle, |$$-$$\alpha\rangle\}$ with equal prior probabilities.
BPSK provides the largest signal distance within any binary signals
under the average-power constraint.
Our receiver discriminates these signals with the BER lower than
that of the SQL.

The physical process of a near-optimal quantum receiver
was first suggested by Kennedy \cite{Kennedy1973}
consisting of linear optics and photon counting.
Soon after it was extended to the exactly optimal one by Dolinar
\cite{Dolinar1973}
(see also \cite{QDET}) via applying an ultrafast feedback process.
A proof-of-principle of the Dolinar receiver
was recently demonstrated \cite{Cook2007}
where instead of the BPSK, intensity modulated (on-off keyed:OOK)
coherent signals $\{|0\rangle, |\alpha\rangle\}$ were discriminated
under the SNL.
However, even if one could achieve the minimum error bound for an OOK
signal discrimination, its error rate is still larger than
the SQL for BPSK signals
since the OOK is not an optimal modulation under
the same power constraint
(Fig.~\ref{fig:schematic}(a)).
Moreover, although in principle the feedback (or feedforward) based
measurement can realize
an arbitrary binary projection measurement \cite{Takeoka2006},
it is still challenging to outperform the SQL for BPSK signals
via the feedback approach
since it requires a detector simultaneously fulfilling
a very high detection efficiency and the operation speed
faster enough than the optical pulse width.
Instead of the feedback, the optimal detection with an optical nonlinear
process was proposed \cite{Sasaki1996}, but
the required nonlinearity was unfortunately far from the current technology.

Instead, we use
a simpler receiver scheme without feedback such as
Kennedy's near-optimal receiver \cite{Kennedy1973}.
The main obstacle to beat the SQL with the Kennedy receiver in practice
is the fact that for weak signals, its attainable BER is comparable
or even higher than that of the SQL \cite{Takeoka2008}.
The receiver scheme employed here is
based on the recent proposal \cite{Takeoka2008},
which we call an optimal displacement receiver (ODR),
solving the above problem and its operation principle was
demonstrated experimentally \cite{Wittmann2008,Tsujino2010}
though their performances could not reach the ideal homodyne performance
due to technical problems, mainly the low detection efficiency
($\lesssim 70$\%) \cite{Wittmann2008,Tsujino2010}
and the low phase stability \cite{Tsujino2010}.

The experiment demonstrated here employs the ODR scheme and
achieves more than 90\% of the total detection efficiency
as well as enough phase stability, mode matching and low dark counts.
To our knowledge, this is the highest total efficiency
demonstrated in any photon-detection based quantum information (and
single-photon level optical communication) protocols.
Such a performance is realized by installing an efficient photon number
resolving detector known as superconducting transition-edge sensor (TES)
\cite{Lita2008,Fukuda2011}),
almost error-free linear optics, and
the TES-based phase stabilization with quantum level signals.

The SQL of the BPSK signal with the average power
$|\alpha|^2$ is expressed as the BER of
$P_{\rm SQL}=(1- {\rm erf}[\sqrt{2}\alpha])/2$
which is simply attained via a perfect homodyne detector.
The optimal measurement achieving the quantum limit, on the other hand,
is described by the projection onto
the quantum superposition of signal states,
$|\pi_-\rangle = b_0 |\alpha\rangle + b_1 |$$-$$\alpha\rangle$
and its complement in the signal space $|\pi_+\rangle$
\cite{QDET}.
Here, $b_0 = -\sqrt{P_{\rm QL}/(1-e^{-4\alpha^2})}$ and
$b_1 = \sqrt{(1-P_{\rm QL})/(1-e^{-4\alpha^2})}$, and
$P_{\rm QL} = (1-\sqrt{1-e^{-4\alpha^2}})/2$ is the fundamental
quantum limit of the BER.
The superposition of coherent states is quite non-classical and
is often regarded as optical ``Schr\"{o}dinger's cat state'' \cite{Yurke1986},
which implies
the implementation of such a projector is a non-trivial task.
The BERs $P_{\rm SQL}$ and $P_{\rm QL}$ are compared
in Fig.~\ref{fig:schematic}(a) showing the high potential of
the optimal quantum measurement.

We implement such a measurement approximately by a simple quantum receiver
\cite{Takeoka2008,Wittmann2008}.
Suppose that coherent signals from the sender are attenuated
in a lossy channel to be $|{\pm}\alpha\rangle$ and detected via the receiver.
Our receiver consists of a linear displacement operation
$\hat{D}(\beta) = \exp[\beta^* \hat{a} - \beta \hat{a}^\dagger]$
and a photon counting device announcing two outcomes, zero
or non-zero photons.
Depending on these outcomes, the signal state is projected onto
$|\omega_-\rangle = \hat{D}^\dagger (\beta) |0\rangle = |$$-$$\beta\rangle$
or its orthogonal space.
Here $\beta$ is optimized such that
$|$$-$$\beta\rangle = e^{-|\beta|^2/2} (|0\rangle - \beta|1\rangle + \cdots)$
approximates the non-trivial superposition
$|\pi_-\rangle = e^{-\alpha^2/2} \{ (b_0-b_1) |0\rangle +
(b_0+b_1) \alpha |1\rangle + \cdots \}$.
In the ideal case, the optimal $\beta$ is given
as the solution of the equality
$\alpha = \beta \tanh (2\alpha\beta)$.
This ODR
outperforms the SQL for any $|\alpha|^2$
and nearly reaches the quantum limit
as illustrated in Fig.~\ref{fig:schematic}(a) \cite{Takeoka2008,Wittmann2008}.

Figure~\ref{fig:schematic}(b) shows an experimental setup for
the optical communication in a lossy channel with
the ODR.
Continuous wave light from an external-cavity laser diode
(wavelength: 853~nm, linewidth: 300~kHz) is highly attenuated and modulated by
an electro-optic modulator (EOM) to generate 20ns pulses with
a 40~kHz repetition rate.
The pulses are split into two paths for the signal
and the auxiliary oscillator (AO) for the displacement operation.
At the sender's station,
the signal amplitude control and
the binary phase modulation are performed by two EOMs.
The signal pulses are then propagated through a channel with
the loss of ${-}7$~dB
where the loss is introduced by coupling the signal beam into
a single-mode fiber with a low efficiency.
The displacement operation $\hat{D}(\beta)$ at the detection part
is realized by interfering the signal pulse
with the AO pulse through a highly transmissive
fiber beam splitter (FBS).
The AO for the displacement is prepared to be
a coherent state $|\beta/\sqrt{1-T}\rangle$
where $T$ is the transmittance of the FBS and chosen as $T>0.99$.
The visibility at the FBS is measured to be $98.6 \pm 0.1\%$
which corresponds to the mode match factor of $\xi=0.993 \pm 0.001$.
The output from the FBS is guided into a TES
to detect photons.


The superconducting TES enables us to
resolve photon numbers with a very high detection efficiency.
Its performance is characterized by
the quantum efficiency and the energy resolution.
Our TES is based on a titanium (Ti) superconductor and
its quantum efficiency is $0.95\pm0.01$ at 853~nm.
This is achieved
by carefully designing the surface (anti-reflection) and backside
(high-reflection) coatings and using a larger Ti device
($10 \times 10$~$\mu$m$^2$) which is coupled to the fiber
with an almost unit efficiency \cite{Fukuda2011}.

The energy resolution characterizes the photon number resolution
and is degraded by the electrical noise on the readout
voltage pulse derived from Johnson noise and phonon noise in TES.
To increase the resolution,
the noise in the output voltage waveform is reduced
via a digital Wiener filter,
which gives an energy resolution of 0.55 eV at 853 nm.
The threshold discriminating the peaks of zero and
non-zero photons is chosen in advance to minimize the
incorrect guessing of photons due to the finite overlap of
the energy distributions.
The additional loss due to the overlap of the distributions
was estimated to be 0.7\%. We also directly observed the dark counts
with the same threshold as $\nu = 0.003$/pulse.

The relative path length of the signal and the AO modes
is actively stabilized by the signal laser and the TES
before measuring BERs.
To circumvent additional losses stemming from the additional probe beam and
optics, the phase locking loop is constructed by using the signal laser
itself ($|\alpha|^2\approx2$) and photon counting
via the TES with a digital feedback.
At the phase locking step, The TES output is sent to a discriminator
and the number of clicks are counted for each 10~ms for the feedback loop.
The stability is estimated to be within $\pm 0.057$ rad.
As a consequence,
after including the optical losses before the TES
(losses at the FBS and fiber splicings),
the total detection efficiency in our setup amounts to $0.91\pm0.01$.

\begin{figure}[h]
\begin{center}
 \includegraphics[width=1\linewidth]{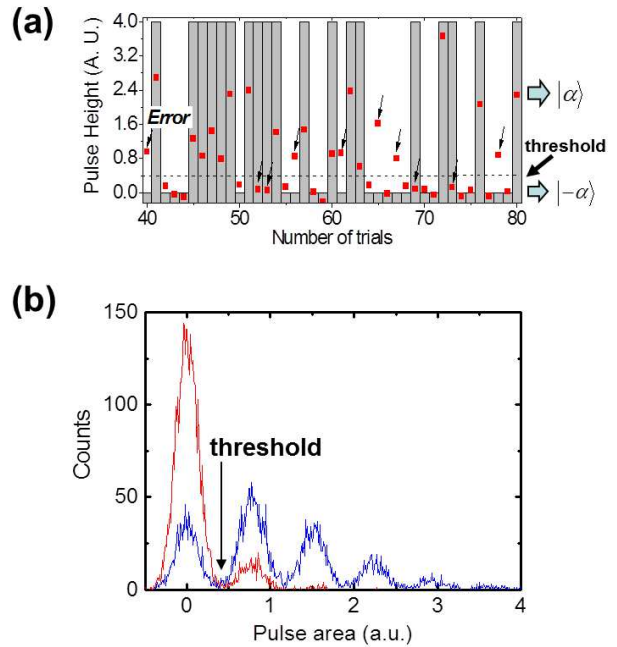}
\end{center}
 \caption{ (color online).
 (a) Random signal modulation (bar) and raw measurement data
of the ODR (plot is the TES output pulse height).
 (b) The output pulse height distribution
for the signals $|{-}\alpha\rangle$ (red)
and $|\alpha\rangle$ (blue).
$|\alpha|^2=0.21$ and $|\beta|^2=0.59$.
}
 \label{fig:RawData}
\end{figure}

\begin{figure}[h]
\begin{center}
 \includegraphics[width=1\linewidth]{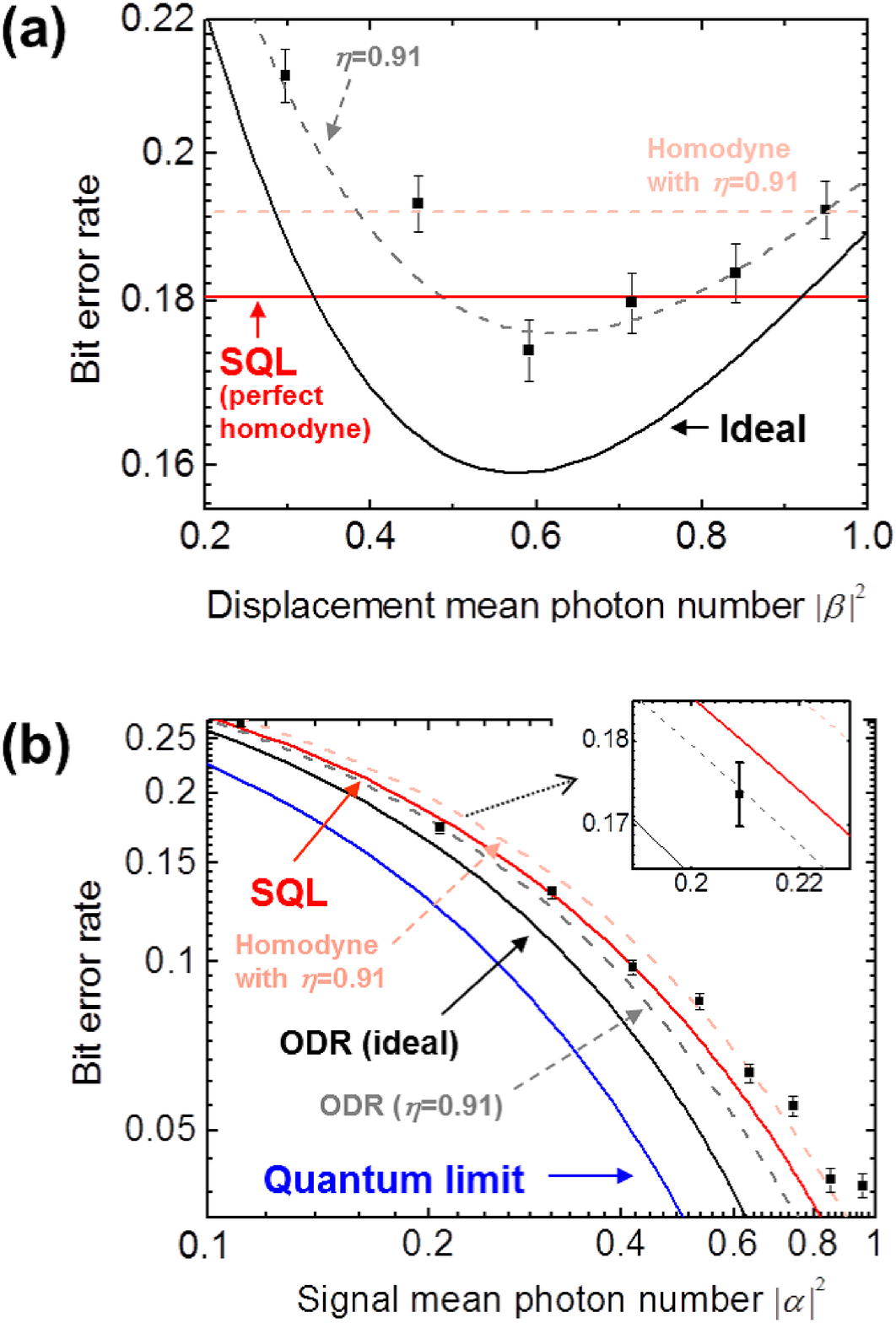}
\end{center}
 \caption{ (color online).
 (a) The measured BERs for various displacements.
$|\alpha|^2$ is fixed to be 0.21.
 (b) The measured BERs obtained by the optimal displacement receiver
for various $|\alpha|^2$.
Theoretical curves are given for the quantum limit (blue solid),
the SQL (red solid), the SQL (homodyne) with $\eta=0.91$ (pink dashed),
and the imperfect displacement receiver (gray dashed).
The imperfections in the gray dashed lines are chosen as
$\eta=0.91$, $\nu=0.003$, and $\xi=0.993$ (see the text for details).
 }
 \label{fig:BER}
\end{figure}


Figures \ref{fig:RawData} and \ref{fig:BER} show
the main result of the BPSK signal discrimination via our quantum receiver.
The phase of the signal is randomly prepared by a pseudo random
number generator with equal probabilities
while the amplitude of the displacement $\beta$ is kept in-phase
with $|\alpha\rangle$.
Figure~\ref{fig:RawData}(a) shows typical sequences of the signal modulation
and the TES output. The dashed line indicates a threshold for the signal
decision and the errors are highlighted by arrows.
The TES output distribution with 10,000 trials is plotted
in Fig.~\ref{fig:RawData}(b) for $|\alpha|^2=0.21$ and $|\beta|^2=0.59$.
The red and blue plots are the outputs for $|{-}\alpha\rangle$
and $|\alpha\rangle$, respectively.
The asymmetric distribution clearly indicates the effect of
the displacement since the original signals $|\pm\alpha\rangle$
have the same photon number statistics.

In Fig.~\ref{fig:BER}(a),
the $\beta$-dependence of BERs is shown
for the signals with a mean photon number of $|\alpha|^2=0.21$.
Each point is obtained by 10,000 measurements
with error bars reflecting statistically estimated standard deviations of
binomial distributions.
For $|\beta|^2 > |\alpha|^2$,
the probability of wrongly guessing $|\alpha\rangle$ ($|{-}\alpha\rangle$)
is decreased (increased) by increasing $|\beta|^2$ and thus
there is an optimal point minimizing the BER (the average errors)
\cite{Tsujino2010}.
The experimental BER shows the dependence on $\beta$ and
clearly outperforms
the SQL at the optimal $\beta$ (around 0.6 photons).
Note that the data points do not employ any compensations
of noises or detection losses.
The experimental BERs make a good fit to the theoretical curve
including imperfections of $\eta=0.91$, $\nu=0.003$,
and $\xi=0.993$ (dashed line) \cite{Takeoka2008}.
The experimental points are also compared to the SQL
(homodyne detection) with the same detection efficiency ($\eta=0.91$),
showing the superiority of our receiver scheme.
In Fig.~\ref{fig:BER}(b),
experimentally observed BERs with optimal displacements are plotted
for different signal photon numbers.
Again, the performance of our receiver clearly outperforms
the SQL ($\eta=1$) at $|\alpha|^2=0.21$,
and is comparable or slightly better than the SQL for the signals
with $|\alpha|^2 \le 0.4$.
For higher signal photon numbers, the BERs deviate from
the model calculation due to technical reasons, mainly
the visibility degradation by the drift of polarizations in fibers.
We should also note that the data shown here outperforms
homodyne detection with the same efficiency ($\eta=0.91$)
in a wide range of signal photon numbers.


In summary,
our result shows the first unconditional observation of the error rate
surpassing the SQL in coherent communication,
that is, exceeding
the theoretical limit of current optical communication technology.
This will open up a new way of extending a link distance in
amplification-free channels such as deep-space optical links,
as well as reducing the number of amplifiers in long-haul
optical fiber communication.
Our scheme could be extended to multiple modulation signals
\cite{Bondurant1993} or another modulation format \cite{Guha2011}.
While the receiver demonstrated here measures each signal separately,
an important future work is to extend it
to the quantum collective decoding \cite{Sasaki1997,Buck2002,Fujiwara2003}
which collectively detects multiple pulses
and is in principle reachable to the ultimate capacity bound
in lossy optical channels \cite{Giovannetti2004}.
From a technical point of view, the photon-level
phase locking implemented here is an important step
toward real communication and could also be useful for
interferometer-based sensing applications.

Finally, it should be noted that our receiver has an ability
to resolve photon numbers.
The displacement receiver with a very high efficiency and
number resolving ability realized here is directly applicable
to quantum information science and technologies such as
a loophole-free test of quantum mechanics with continuous variable
states \cite{Banaszek1999},
an efficient receiver for coherent state-based quantum cryptography
\cite{Wittmann2010}, and quantum repeaters and computation based on
entangled coherent states \cite{Ralph2003,Spiller2006,Azuma2010}.

{\bf Acknowledgements.}
The authors thank H. Takahashi, E. Sasaki, and T. Itatani for their
technical supports and J. Neergaard-Nielsen for reading
of the manuscript. This work has been supported by
MEXT Grant-in-Aid
for Young Scientists (B) 22740720.


\providecommand{\noopsort}[1]{}\providecommand{\singleletter}[1]{#1}%

\end{document}